# NR Wide Bandwidth Operations

Jeongho Jeon, Intel Corporation

*Abstract*—The 3rd Generation Partnership Project (3GPP) is in the process of developing the next generation radio access technology, named New Radio (NR), which will be proposed as a candidate technology for IMT-2020. This article outlines the wide bandwidth operation of NR, among other new features being considered, based on the up-to-date discussions and decisions made in 3GPP standardization meetings. The much wider channel bandwidth of NR, compared to LTE, enables more efficient use of resources than the existing carrier aggregation framework at lower control overhead. The support of multiple sub-carrier spacing options allows NR to operate in a wide range of carrier frequency from sub-6 GHz band to mmWave band with appropriate handling of multi-path delay spread and phase noise depending on the carrier frequency. In addition, the introduction of the new bandwidth part concept allows to flexibly and dynamically configure User Equipment's (UE's) operating bandwidth, which will make NR an energy efficient solution despite the support of wide bandwidth. Other NR wideband operation related issues, such as the support of UEs with limited radio frequency (RF) capability and frequency domain resource indexing, are also explained in this article.

## I. INTRODUCTION

New technologies such as automated intelligence, the Internet of Things (IoT), autonomous vehicles, virtual/augmented reality, etc., are becoming reality. These technologies are based on even faster, more prompt, and more reliable inter-connectivity of everything, which demands for the next generation of mobile communication systems. In accordance with such demands, the International Telecommunication Union Radiocommunication Sector (ITU-R) suggested the recommendation to International Mobile Telecommunications (IMT) for 2020 and beyond, i.e., IMT-2020 [1]. The recommendation enumerates key envisaged usage scenarios for IMT-2020 including enhanced mobile broadband (eMBB), ultra-reliable low latency communications (URLLC), and massive machine type communications (mMTC). The *Guidelines for Evaluation of Radio Interface Technologies for IMT-2020*, in its draft form, describes the procedure, methodology and the criteria to be used in evaluating the candidate IMT-2020 radio interface technologies. Candidate technologies to be submitted[1] to ITU-R will be finally accepted for inclusion in the IMT-2020 radio interface recommendation(s), if it is determined by ITU-R that the technology fulfils the minimum technical performance requirements. The evaluation criteria of the IMT-2020 contains the following performance indices:

- *Peak data rate, peak spectral efficiency, user experienced data rate, 5th percentile user spectral efficiency, average spectral efficiency, area traffic capacity, user plane latency, control plane latency, connection density, energy efficiency, reliability, mobility, mobility interruption time, bandwidth, support of wide range of services, and supported spectrum band(s)/range(s).*

New Radio (NR), which is currently being developed by the 3rd Generation Partnership Project (3GPP), is a candidate technology for potential inclusion in the IMT-2020 radio interface recommendation(s) by ITU-R. 3GPP has been the development and/or maintenance hub of Global Systems for Mobile Communications (GSM) and related 2/2.5G standards, Universal Mobile

---

[1] The final deadline for submissions is prior to the start of the 32nd meeting of ITU-R Working Party (WP) 5D in July 2019.

Telecommunications System (UMTS) and related 3G standards, Long Term Evolution (LTE) and related 4G standards, and NR and related 5G standards. There are three Technical Specification Groups (TSGs), namely Radio Access Network (RAN), Service and Systems Aspects (SA), and Core Network and Terminals (CT). Each TSG consists of multiple Working Groups (WGs)[2].

TSG RAN had conducted pre-standardization research on NR during its Study Item (SI) phase in 3GPP Release-14 from April 2016 to March 2017 [2]. TSG RAN is currently conducting the NR Phase I standardization in its Work Item (WI) phase in 3GPP Release-15 until June 2018, which will be followed by the NR Phase II standardization in 3GPP Release-16 until December 2019. The RAN1 WG is currently seeking improvements in the physical layer from all aspects to fulfil the IMT-2020 performance requirements and, thereby, to successfully support the main usage scenarios suggested by ITU-R. For instance, RAN1 is specifying the support of multi-panel multi-input multi-output (MIMO) antenna systems from various aspects, e.g., new codebook design for hybrid analog/digital beamforming and improved channel state information (CSI) estimation and reporting mechanisms, etc., which is essential to increase the data rate and spectral efficiency. The support of new coding schemes, such as the low-density parity-check (LDPC) code for the data channel and the polar code for the control channel (at least for eMBB in the current status), aims to improve the data rate and reliability and reduce the latency. Additionally, many other new technical components, including dynamic TDD, self-contained frame structure, various improvements in downlink (DL)/uplink (UL) control mechanisms, and the support of NR-LTE coexistence, are currently being developed.

This article outlines the NR wide bandwidth (BW) operations based on the up-to-date discussions and decisions made in 3GPP standardization meetings among other new features being considered to meet the IMT-2020 requirements. More specifically, NR supports scalable sub-carrier spacing (SCS) and much wider channel BW (CBW) compared to LTE, which are directly related to *data rate, latency, bandwidth, support of a wide range of services,* and *supported spectrum band(s)/ranges(s)* amongst the aforementioned IMT-2020 requirements. The wide CBW enables more efficient use of resources than the existing carrier aggregation (CA) mechanism. On the other hand, there are several advantages of supporting multiple SCS options in a system. With smaller SCS, a system can be made more tolerable to the effect of multi-path delay spread while keeping the ratio of the cyclic prefix (CP) duration to the orthogonal frequency-division multiplexing (OFDM) symbol duration, i.e., CP overhead ratio, unchanged. This is especially useful for lower carrier frequency. With larger SCS, it is easier to estimate and compensate the phase noise, which is aggravated in higher carrier frequency. More details will be described in Section II. In supporting the wide CBW, NR design considers the situation when there are User Equipments (UEs) with limited radio frequency (RF) capability such that they cannot cover a wideband carrier from a network perspective with single RF chain. To this end, NR allows to simultaneously operate some high capable UEs with one wideband carrier, while at the same time, some low capable UEs via intra-band contiguous CA within the wideband carrier from a network perspective. Consequently, the notion of a carrier became rather UE-specific in NR than cell-specific as in LTE. This is in consideration of challenges during early-stage NR UE implementation. Details will be further discussed in Section III. Furthermore, NR provides a mechanism to adaptively adjust UEs' operating BW via the introduction of the bandwidth part (BWP) concept, where a UE is not required to transmit or receive outside of the configured frequency range of the active BWP with an exception of measurement gap. The BWP is a crucial component of NR system to improve the *energy efficiency*, which is one of the IMT-2020 requirements. The BWP concept will be described in details in Section IV. Lastly, a new frequency domain resource indexing scheme for NR is explained in accordance with the introduction of wide CBW and BWP concept.

This article aims to help in understanding the overall NR wideband operations. Throughout the article, unique features of NR,

---

[2] For instance, TSG RAN consists of RAN1 for the specification of the physical layer of the radio interface between the UE and the network, RAN2 for the specification of the radio interface architecture and protocols, RAN3 for the specification of the overall network architecture, RAN4 for the specification of the minimum performance requirements, and RAN5 for the specification of the UE conformance testing.

regarding the wideband operations, will be explained in comparison with LTE, along with the motivation for such introduction.

## II. NR Sub-carrier Spacing and Channel Bandwidth

### A. Sub-carrier Spacing

NR supports scalable SCS compared to LTE's single 15 kHz SCS option[3]. The scalable SCS is expressed as $f_c = 15 \cdot 2^n$ [kHz], where $n$ is a non-negative integer. The power-of-two scaling law facilitates the symbol boundary alignment between different SCSs. In other words, the symbol at some SCS $f_1$ always exactly contains an integer number of symbols at any SCS $f_2$, which is larger than $f_1$. RAN1 has decided to support up to 480 kHz SCS from RAN1 specification point of view during the SI phase [2], but the actual set of SCS values that can be used in each band is up to RAN4. According to the RAN4's decision [3], in Release-15, 15/30/60 kHz SCS values are used in frequency range 1 (FR1) with up to 100 MHz CBW, where FR1 denotes the frequency range of 450 – 6000 MHz, aka sub-6 GHz bands. 60/120 kHz SCS values are used in frequency range 2 (FR2) with up to 400 MHz CBW, where FR2 denotes the frequency range of 24.25 – 52.6 GHz, aka mmWave bands. The supported SCS and the CBW values vary from one band to another, which will be listed in Section 4.2 *Channel bandwidth* of 3GPP Technical Report (TR) 38.817-01 *General Aspects for UE RF for NR* [4]. Note that there could be exceptions for non-data channels. For instance, in NR, the primary and secondary synchronization signals (PSS/SSS) and the Physical Broadcast Channel (PBCH), which are collectively termed as SS block, will use 15/30 kHz SCS for sub-6 GHz and 120/240 kHz for above-6 GHz frequency bands. In the case of the Physical Random Access Channel (PRACH), the long preamble sequence utilizes 1.25/5 kHz SCS, in addition to the short preamble sequence using 15/30/60/120 kHz SCS.

There are several advantages of supporting multiple SCS options in NR. When smaller SCS is used, the symbol length increase is inversely proportional. With longer symbol length, the CP of the OFDM symbol can be longer at the same ratio of the CP duration to OFDM symbol duration, i.e., the CP overhead ratio. Thus, with smaller SCS, a system can be made more tolerable to the effect of multi-path delay spread at the same CP overhead ratio. On the other hand, larger SCS makes a symbol duration shorter, which is not only beneficial for fast transmission turnaround time but also advantageous in terms of lower sensitivity to phase noise. The phase noise is a random process and directly impacts up/down conversion between baseband and RF signals due to the temporal instability of the local oscillator(s). The phase noise in the frequency domain gives rise to signal jitter in the time domain. In general, the phase noise increases with carrier frequency and, therefore, it is more problematic in higher carrier frequency. When the rate of phase variation is slow, with respect to the OFDM symbol duration, the phase noise can be modeled as a constant and can be compensated via estimation. The reference signals (RSs) in NR for such purpose are phase tracking RSs (PT-RSs) and fine time/frequency tracking RSs. However, when the rate of phase change is faster with respect to the OFDM symbol duration, the estimation of phase noise and, thereby, the correction becomes difficult. Therefore, the larger the SCS, the easier it is to compensate the phase noise. On the other hand, as the carrier frequency goes higher, e.g., mmWave band, the signal propagation exhibits less multi-path delay spread due to sharp beamforming with massive MIMO antenna and the signal propagation characteristics itself at higher frequency. Thus, having longer CP becomes less important in higher carrier frequency. This is one of the reasons why only 60/120 kHz SCS options are supported in high frequency bands.

### B. Channel Bandwidth

NR system supports much wider maximum CBW than LTE's 20 MHz. Wideband communication is also supported in LTE via CA of up to 20 MHz component carriers (CCs). By defining wider CBW in NR, it is possible to dynamically allocate frequency resources via scheduling, which can be more efficient and flexible than the CA operation, whose activation/deactivation is based on the medium access control (MAC) control element (CE). The activation/deactivation via MAC CE is regarded a reliable option as it is protected by hybrid automatic repeat request (HARQ), but the reliability comes at the cost of longer latency for activation/deactivation due to the HARQ feedback procedure. Having single wideband carrier also has merit in terms of low control overhead as it needs only single control signaling, whereas CA requires separate control signaling per each aggregated carrier. In this regard, the maximum CBW per NR carrier, which was agreed in RAN1 during the NR SI phase, is 400 MHz in Release-15 from a physical layer specification perspective [2]. As it was described in Section II-A, it has been decided by RAN4 to support up to 100 MHz CBW in FR1 and up to 400 MHZ CWB in FR2 in Release-15 [3]. Actual supported CBW values are dependent not only on the specific band but also on the used SCS due to the limitation on the fast Fourier transform (FFT) size [4].

---

[3] The LTE DL/UL data and control channels use 15 kHz SCS, but smaller than 15 kHz SCS values are also used e.g., for Physical Multicast Channel (PMCH), Physical Random Access Channel (PRACH), and Narrowband Physical Uplink Shared Channel (NPUSCH) for Narrowband IoT (NB-IoT).

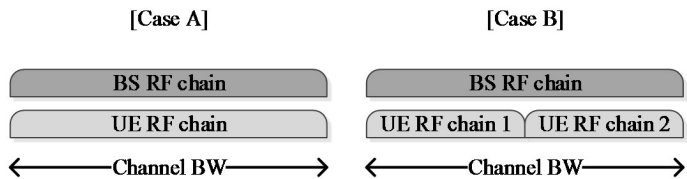

Fig. 1. RF configurations of UE and base station for a given NR channel BW

Like LTE, NR also supports the aggregation of multiple carriers via CA or dual connectivity[4] (DC). Note that LTE supports up to 32 CCs thanks to the Release-13 LTE CA enhancement WI. It has been decided for NR to support up to 16 CCs, which is less than LTE's maximum aggregation. Note, however, that in the case of NR, the CBW itself is much wider than LTE and, thus, the motivation for supporting very large number of CCs is not significant.

*C. FFT Size*

During the NR SI phase, RAN1 has decided that the candidates for the maximum number of subcarriers per NR carrier are 3,300 and 6,600 in Release-15 from a physical layer specification perspective [2]. Assuming FFT implementation using radix-2 algorithm, these candidates of the maximum number of subcarriers imply 4K and 8K FFT sizes, respectively. The 4K FFT has advantages over 8K FFT in terms of ease of implementation, FFT block die size, and power consumption. On the other hand, 8K FFT can allow a smaller SCS value for a given CBW than 4K FFT. This can be beneficial for wide CBW, which is available in mmWave band. However, as discussed in Section II-A, the need for smaller SCS is weakened in mmWave band due to the smaller multi-path delay spread. Consequently, none of the maximum CBW and the supported SCS pairs defined by RAN4 require more than 3,300 sub-carriers in a carrier [4]. Therefore, the maximum number of sub-carriers per NR carrier is 3,300, and the FFT size is 4K for NR Phase I in Release-15.

III. UE WITH LIMITED RF CAPABILITY

By defining wide CBW for NR, there is a need to consider UEs having limited RF capabilities such that the UE cannot cover a wideband carrier from a network perspective with a single RF chain. Such consideration is more important in the early-stage NR deployment. For instance, as depicted in Figure 1, there could be UEs that can cover the wide CBW with a single RF chain (Case A), whereas there could be UEs that can cover the wide CBW only with more than one chains (Case B). In other words, the NR system needs to support the situation where UEs with different RF capabilities are coexisting in the same network.

---

[4] CA and DC are similar in terms of aggregating multiple carriers. The difference is the applicable deployment scenario; CA is for scenarios with ideal backhaul and DC is for scenarios with non-ideal backhaul, e.g., non-collocated base stations with communication latency between them.

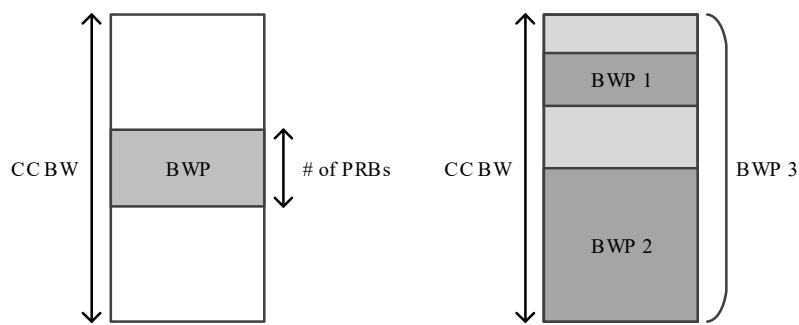

Figure 2. An example of BWP configurations

In order to handle the coexistence of UEs having different RF capabilities, the following two alternatives were discussed:
- Alternative 1: The network configures both Case A and Case B UEs with one wideband CC, while Case B UEs utilize multiple RF chains to cover one wideband CC.
- Alternative 2: The network configures Case A UEs with one wideband CC, while at the same time the network configures Case B UEs with a set of intra-band contiguous CCs with CA.

When a signal is transmitted over multiple RF chains, the signal will experience amplitude and phase discontinuity between the RF chains. Therefore, the channel estimation and the signal transmission need to be performed separately over each RF chain. Accordingly, it was agreed to adopt Alternative 2 for NR.

Note that such flexible network operation tailored to each UE's different RF capability, i.e., simultaneously operating as wideband CC for some UEs and, at the same time, as intra-band CA for some other UEs, is not found in the LTE system. Consequently, the notion of CC in NR is rather UE-specific than cell-specific as in LTE. Although it was not depicted in Figure 1, there could be some other UEs which only have single limited RF chains and, thus, covering only part of the wideband CC from a network perspective; with the UE-specific notion of NR CC, such UEs can also be naturally accommodated. Lastly, from the view point of efficiently utilizing the frequency resources, it is beneficial to minimize the guard-band, preferably to zero, between CCs for Case B UEs within wideband carrier.

## IV. BANDWIDTH PARTS

*A. Background*

Wide BW has direct impact on the peak and user experienced data rates. However, since UEs are not always demanding high data rates, the use of wide BW may imply higher idling power consumption both from RF and baseband signal processing perspectives. In this regard, a newly developed concept of BWP for NR provides a means of operating UEs with smaller BW than the configured CBW, which makes NR an energy efficient solution despite the support of wideband operation. Alternatively, one may consider to schedule a UE such that it only transmits or receives within a certain frequency range. Compared to this approach, the difference with BWP is that the UE is not *required* to transmit or receive outside of the configured frequency range of the active BWP, which attributes power saving from the following aspects [5]:
- *There may be power savings in some scenarios due to the possibility to operate the RF-baseband interface with a lower sampling rate and reduced baseband processing needed to transmit or receive with narrower bandwidth.*
- *UE RF bandwidth adaptation can provide UE power saving at least if carrier bandwidth before adaptation is large.*

One note here is that as the UE power consumption is quite dependent on each modem and RF implementation, it is difficult to expect a quantitative power saving gain.

*B. Configuration*

As depicted in Figure 2, a BWP consists of a group of contiguous physical resource blocks (PRBs). The BW of a BWP cannot exceed the configured CC BW for the UE. The BW of the BWP must be at least as large as one SS block BW, but the BWP may or may not contain SS block. Each BWP is associated with a specific numerology, i.e., SCS and CP type. Therefore, the BWP is also a means to reconfigure a UE with a certain numerology. As illustrated in the right figure of Figure 2, the network can configure multiple BWPs to a UE via Radio Resource Control (RRC) signaling, which may overlap in frequency. The granularity of BW configuration is one PRB. For each serving cell, DL and UL BWPs are configured separately and independently for paired spectrum and up to four BWPs can be configured for DL and UL each. For unpaired spectrum, a DL BWP and a UL BWP are jointly configured as a pair and up to 4 pairs can be configured. There can be maximally 4 UL BWPs configured for a supplemental UL (SUL) as well[5].

Each configured DL BWP includes at least one control resource set (CORESET) with UE-specific search space (USS). The USS is a searching space for UE to monitor possible reception of control information destined for the UE. In the primary carrier, at least one of the configured DL BWPs includes one CORESET with common search space (CSS). The CSS is a searching space for UE to monitor possible reception of control information common for all UEs or destined for the particular UE. If the CORESET of an active DL BWP is not configured with CSS, the UE is not required to monitor it. Note that UEs are expected to receive and transmit only within the frequency range configured for the active BWPs with the associated numerologies. However, there are exceptions; a UE may perform Radio Resource Management (RRM) measurement or transmit sounding reference signal (SRS) outside of its active BWP via measurement gap. The BWP is also a tool to switch the operating numerology of a UE. The numerology of the DL BWP configuration is used at least for the Physical Downlink Control Channel (PDCCH), Physical Downlink Shared Channel (PDSCH) and corresponding demodulation RS (DMRS). Likewise, the numerology of the UL BWP configuration is used at least for the Physical Uplink Control Channel (PUCCH), Physical Uplink Shared Channel (PUSCH) and corresponding DMRS. On the other hand, it is noted that there is a restriction in the configuration of numerology at least in the early version of NR. That is, the same numerology shall be used within the same PUCCH group including both DL and UL[6].

*C. Activation/Deactivation*

Multiple options could be supported for activation/deactivation of BWPs. In addition to the activation/deactivation via dedicated RRC signaling, downlink control information (DCI) based activation/deactivation is supported. On the other hand, the DCI based mechanism, although more prompt than the one based on MAC CE, requires additional consideration for error case handling, i.e., the case when a UE fails to decode the DCI containing the BWP activation/deactivation command. To help to recover from such a DCI lost case, the activation/deactivation of DL BWP (or DL/UL BWP pair for the case of unpaired spectrum) by means of timer is also introduced. With this mechanism, if a UE is not scheduled for a certain amount of time, i.e., expiration of timer, the UE switches its active DL BWP (or DL/UL BWP pair) to the default one. There is an initial active BWP for a UE during the initial access until the UE is explicitly configured with BWPs during or after RRC connection establishment. The initial active BWP is the default BWP, unless configured otherwise. In Release 15, for a UE, there is at most one active DL BWP and at most one active UL BWP. The HARQ retransmission across different BWPs is supported when a UE's active BWP is switched.

---

[5] SUL is introduced for NR mainly to compensate the NR UL coverage.
[6] The only exception of this restriction is with the SUL, which can have a smaller SCS than the associated DL carrier.

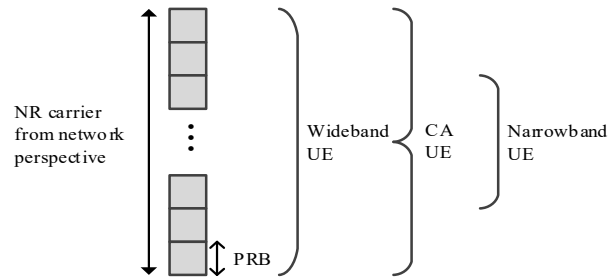

Figure 3. Coexistence of wideband, carrier aggregation, and narrowband UEs

*D. RF Retuning Time*

When it is expected that a UE performs RF adaptation due to BWP switching, physical RF retuning time needs to be taken into account. The following is RAN4's preliminary understanding on the RF retuning time [5]:

– *For intra-band operation, at least for sub-6 GHz, the transition time can be up to 20 μs if the center frequency is the same before and after the bandwidth adaptation.*
– *For intra-band operation, at least for sub-6 GHz, the transition time is 50~200 μs if the center frequency is different before and after the bandwidth adaptation.*
– *For inter-band operation, at least for sub-6 GHz, the transition time can be up to 900 μs.*

Since BWPs are configured within an NR carrier, the inter-band RF retuning time in the last bullet point above is not relevant to the BWP switching. It is also noted that the RF retuning time for mmWave needs to be studied as well. For unpaired spectrum, the UE is not expected to retune the center frequency of the CBW between DL and UL. Considering the RF retuning time for BWP switching, if a PDCCH and the corresponding PDSCH scheduled by the PDCCH are transmitted in different BWPs, a UE does not expect the PDSCH to start earlier than $K$ symbols after the end of the corresponding PDCCH. The active BWP switching time is up to RAN4 including whether the transition time(s) of active BWP switching is reported to the network as UE capability or not.

V. PHYSICAL RESOURCE BLOCK INDEXING

As described in Section III and also illustrated in Figure 3, in a given wideband NR carrier from a network perspective, wideband UEs, CA UEs, and narrowband UEs, depending on their RF implementation, can coexist. Moreover, even for a given UE, the active BWP can change over time. Therefore, a special handling is necessary for NR. To this end, NR supports both common PRB indexing and UE-specific PRB indexing, which is different from LTE that only needs and supports cell-specific PRB indexing. The common PRB indexing is common to all the UEs sharing a wideband CC from a network perspective regardless of whether they are wideband, CA, or narrowband UEs. In order for a UE to recognize the common PRB indexing, the UE is signaled on the offset between a reference location and the lowest subcarrier of the reference PRB, which is the lowest PRB of the carrier from a network perspective including potential guard bands. The offset in the unit of PRB is indicated based on 15 kHz SCS for FR1 and 60 kHz SCS for FR2. A reference location for DL in the primary cell (PCell), for instance, is the lowest subcarrier of the lowest PRB of the SS block accessed by the UE during the initial access. The expected usage of the common PRB indexing is for scheduling a group common PDSCH, RS sequence generation and reception, and BWP configuration, etc. It is further emphasized that the use of common PRB indexing is of significant importance to correctly identify the portion of the RS sequence within the wideband carrier from a network perspective and to maintain the orthogonality between the RS sequences generated by different UEs. The UE-specific PRB indexing is indexed per BWP with respect to the configured SCS for the configured frequency range of the BWP and, thereby, requires less number of bits to express the resource allocation, which has an impact on the DCI bit-field design. The UE-specific PRB indexing will be used for scheduling a UE-specific PDSCH.

## VI. Conclusion

3GPP is currently developing the NR system, which will be considered as a candidate technology for IMT-2020. Among various improvements that 3GPP is seeking, the wideband operation is one of the key aspects of defining NR. The wide channel bandwidth and multiple SCS options enable the NR system to efficiently and flexibly operate from sub-6 GHz band to mmWave band with appropriate handling of multi-path delay spread and phase noise depending on the carrier frequency. It is also intended to support high data rates for various services with different latency requirements by changing the transmission turnaround time though varying the SCS. Moreover, despite the wide channel bandwidth, the new concept of BWP in NR allows energy efficient UE operation. Last but not least, the NR system makes a practical consideration for initial deployment concerning UEs with different RF capabilities. All of these features are distinguished from the previous generations' radio access technologies.

## Biographies

Jeongho Jeon (jeongho.jeon@intel.com) received the Ph.D. degree in electrical engineering from the University of Maryland, College Park, MD, USA, in 2013. Since then, he has been with Intel Corporation, where he conducts research and standardization of next generation wireless technologies. He is a recipient of the National Institute of Standards and Technology (NIST) Fellowship from 2011 to 2013 and the 14th Samsung Humantech Thesis Prize in 2008.

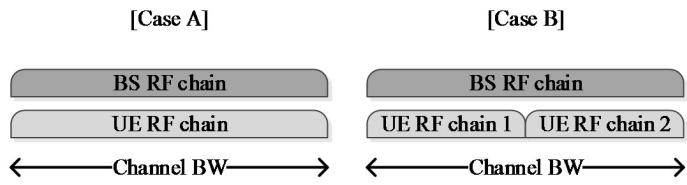

Fig. 1. RF configurations of UE and base station for a given NR channel BW

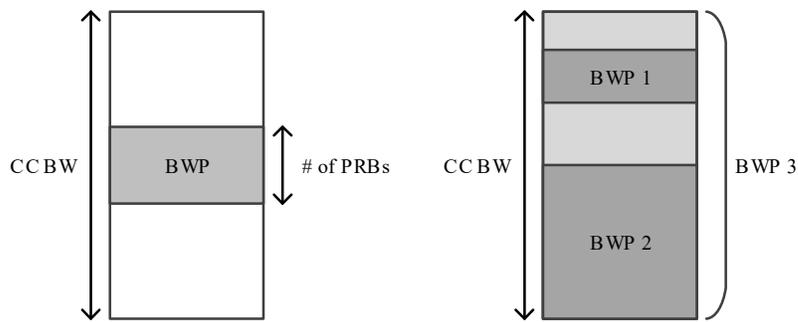

Figure 2. An example of BWP configurations

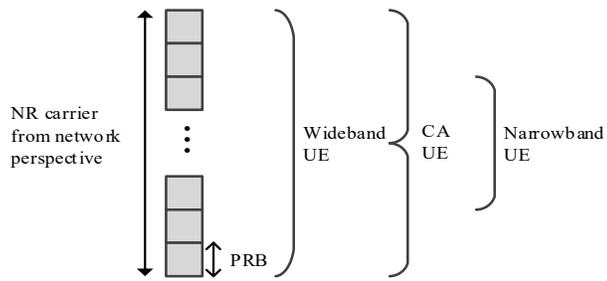

Figure 3. Coexistence of wideband, carrier aggregation, and narrowband UEs